\documentclass[ 12pt]{revtex4}
\usepackage{graphics}

\begin{document}

\newcommand{\n}{\eta}
\newcommand{\nb}{\bar{\eta}}
\newcommand{\np}{\eta^{\prime}}
\newcommand{\nbp}{\bar{\eta}^{\prime}}
\newcommand{\bk}{{\bf k}}
\newcommand{\bl}{{\bf l}}
\newcommand{\bq}{{\bf q}}
\newcommand{\br}{{\bf r}}
\newcommand{\z}{z}
\newcommand{\zb}{\bar{z}}
\newcommand{\zp}{z^{\prime}}
\newcommand{\zbp}{\bar{z}^{\prime}}
\newcommand{\la}{\lambda}
\newcommand{\lab}{\bar{\lambda}}
\newcommand{\w}{\omega}
\newcommand{\e}{\epsilon}
\newcommand{\psib}{\bar{\psi}}
\newcommand{\psip}{\psi^{\prime}}
\newcommand{\psibp}{\bar{\psi}^{\prime}}
\newcommand{\phib}{\bar{\phi}}
\newcommand{\phip}{\phi^{\prime}}
\newcommand{\phibp}{\bar{\phi}^{\prime}}
\newcommand{\g}{\Gamma}
\newcommand{\s}{\sigma}

\title{Non-Markovian Entanglement Dynamics of Two Qubits Interacting
with a Common Electromagnetic Field}
\author{C.~Anastopoulos$^{1}$\footnote{Corresponding author. Email
address: anastop@physics.upatras.gr},
S.~Shresta$^{2,3}$\footnote{Present address: MITRE Corporation 7515
Colshire Drive, MailStop N390 McLean, VA 22102. Email Address:
sanjiv$\_$shresta@mitre.org},  and B.~L. Hu$^{2}$\footnote{Email
address: blhu@umd.edu}} \affiliation{$^1$Department of Physics,
University of Patras, 26500 Patras, Greece, \\$^2$Joint Quantum
Institute, Department of Physics,
University of Maryland, College Park, Maryland 20742-4111 \\
$^3$NIST, Atomic Physics Division, Gaithersburg, MD 20899-8423}

\begin{abstract}
We study the non-equilibrium dynamics of a pair of qubits made of
two-level atoms separated in space with distance $r$ and interacting
with one common electromagnetic field but not directly with each
other. Our calculation makes a weak coupling assumption but no Born
or Markov approximation. We write the evolution equations of the
reduced density matrix of the two-qubit system after integrating out
the electromagnetic field modes. We study two classes of states in
detail: Class A is a one parameter family of states which are the
superposition of the highest energy and lowest energy states, and
Class B states which are the linear combinations of the symmetric
and the antisymmetric Bell states. Our results for an initial Bell
state are similar to those obtained before for the same model
derived under the Born-Markov approximation. However, in the Class A
states the behavior is qualitatively different: under the
non-Markovian evolution we do not see sudden death of quantum
entanglement and subsequent revivals, except when the qubits are
sufficiently far apart. We provide explanations for such differences
of behavior both between these two classes of states and between the
predictions from the Markov and non-Markovian dynamics. We also
study the decoherence of this two-qubit system.
\end{abstract}

\maketitle

\section{Introduction}


Quantum decoherence \cite{PazZurRev} and entanglement \cite{Hor09} between one two-level atom (2LA) and an electromagnetic field (EMF) has been treated by  many authors (see, e.g., the review \cite{FicTan}) including the present ones \cite{AH,SADH,CumHu06}. Zyczkowski et al \cite{Karol} have studied discrete dynamics of entanglement of a bipartite quantum state. 
For two 2LA-EMF systems, each of the two atoms can be assumed to
interact only with its own cavity EMF, or with a common EMF (of
course they can also interact with each other). The case of each
atom interacting with its own field was studied by Yu and Eberly
\cite{YuEbePRL,YuEbePRB} who reported on the appearance of `sudden
death', or finite time disentanglement. The case of two atoms
interacting through a common field was studied by Ficek and Tanas
under the Born-Markov approximation (BMA) \cite{FicTan06}. In
contrast to the `sudden death' found in two qubits in disjoint EMFs
they show the existence of dark periods and revival of quantum
entanglement. The present authors have studied the case of two 2LA
interacting through a common electromagnetic field without invoking
the BMA \cite{ASH}.  This essay is a synopsis of this unpublished
work, focussing on the new non-Markovian behavior of entanglement
dynamics and decoherence features.


In our prior work \cite{AH,SADH}, we used the influence functional
formalism with a Grassmannian algebra for the qubits (system) and a
coherent state path integral representation for the EMF
(environment). Here an operator method is used in conjunction with
perturbation theory. Because the assumption of an initial vacuum
state for the EMF allows a full resummation of the perturbative
series, a closed expression for the evolution of the reduced density
matrix of the two qubits can be obtained. This approach incorporates
the back-action of the environment (field) on the system (two 2LA)
self-consistently which (almost always) engenders non-Markovian
system dynamics the behavior of which cannot be fully comprehended if one imposes the restrictions of BMA.

In this broader context we can understand certain effects like `sudden death' \cite{YuEbePRL} as consequences of rather special arrangements: Each atom interacting with its own EMF precludes or lessens the chance that the atoms may be entangled through the mediation of the EMF. The separate field case studied by Yu and Eberly corresponds to the limit of the common field case studied here when the qubits are separated by distances much greater than the correlation length characterizing the total system. Our study shows that for a wide range of spatial
separations within the correlation length, entanglement is robust and there is no sudden death. The difference between our results and that of Ref. \cite{FicTan06} highlights the non-Markovian effects in the evolution of quantum entanglement. 

Under the usual two level, dipole and rotating wave approximations
(RWA) but without invoking the Born-Markov approximation (BMA) we
derive a non-Markovian master equation which has a more elaborate
structure than the usual Lindblad form: it contains extra terms that
correspond to off-diagonal elements of the density matrix
propagator. We observe very different behavior in two classes of
states, superpositions of highest and lowest energy states which we
call Class A states, and the usual antisymmetric $|- \rangle$ and
symmetric $|+\rangle $  Bell states \cite{Bell}, which we call Class
B states. We find similar behavior in the Class B (Bell) states
compared with the findings of Ref \cite{FicTan06} but qualitatively
different behavior in the evolution of Class A states.  Ficek and
Tannis \cite{FicTan06} found that their evolution leads generically
to sudden death of entanglement and a subsequent revival. In our
more complete treatment of the atom-field dynamics we see the former
effect present for large values of the inter-qubit distances.
However, sudden death is absent for short distances, while there is
no regime in which a revival of entanglement can take place. (Both
our results and the ones of Ref. \cite{FicTan06} rely on the
Rotating Wave Approximation, which is not adequate for very large
interqubit distances -- details in sections below.)

This paper is organized as follows: Section 2 contains the main
derivation. We write down the Hamiltonian for two 2-level atoms
interacting with a common electromagnetic field at zero temperature,
and we compute the relevant matrix elements for the propagator of
the total system by resummation of the perturbative series. We then
determine the evolution of the reduced density matrix of the atoms,
which is expressed in terms of seven functions of time. We compute
these functions using an approximation that amounts to keeping the
contribution of the lowest loop order for the exchange of photons
between the qubits. In Section 3 we examine the evolution of the
reduced density matrix for two classes of initial states. We then
describe the time evolution of quantum entanglement in these states  via the concurrence function
\cite{Wootters} with spatial separation dependence plotted for some representative cases. We discuss
the results on disentanglement, describe the non-Markovian dynamics
and identify the source of difference from quantum dynamics under
the Born-Markov approximation.  In Section 4, we study the
decoherence of this system when the two qubits are initially
disentangled.

In the present synopsis we shall describe our approach, the main steps in the derivations and the physical significance of our findings but leave many details out as they can be found in \cite{ASH}. As a literature update (from the time of \cite{ASH}) on nonMarkovain entanglement dynamics we mention the work of the Turku group \cite{Turku} on atom-field interaction with experimental verification possibilities and that of Paz and Roncaglia \cite{PazRon}, Lin and Hu \cite{LinHu2HOEnt} on two harmonic oscillators (detectors) interacting with a common EMF.  The latter paper addresses a wider scope of issues including outside-light cone entanglement, and entanglement generation from initial separable states with separation dependence.  Lin and Hu find that when two initially entangled detectors are still outside each other's lightcone, the proportionality of the degree of entanglement to the spatial separation oscillates in time. When the two detectors begin to have causal contact, an interference pattern of the relative degree of entanglement (compared to those at spatial infinity) develops a parametric dependence on $d$. The detectors separated at those $d$ with stronger relative degree of entanglement enjoy longer disentanglement times. In the cases with weak coupling and large separation, the detectors always disentangle at late times. For sufficiently small $d$, the two detectors can have residual entanglement even if they initially were in a separable state, while for $d$ a little larger, there could be transient entanglement created by mutual influences.

\section{Two 2-Level Atoms interacting via a common Electromagnetic Field}

\subsection{The Hamiltonian}

We consider two 2-level atoms (2LA),  acting as two qubits, labeled
1 and 2, and an electromagnetic field (EMF) described by the free
Hamiltonian $H_0$
\begin{eqnarray}
\hat{H}_0 = \hbar \sum_a \omega_a \hat{b}^{\dagger}_a \hat{b}_a
+\hbar\omega_o  \hat{S}_+^{(1)} \hat{S}_-^{(1)} +\hbar\omega_o
\hat{S}_+^{(2)} \hat{S}_-^{(2)}
\end{eqnarray}
where $a = (\bk, \sigma)$ is a collective index running over momenta
$\bk$ and polarizations $\sigma = 1, 2$ of the photon bath,
$\omega_a = \omega_\bk$ is the frequency of the $\bk^{\mbox{th}}$
electromagnetic field mode and $\omega_o$ the atomic frequency
between the two levels of the atom, assumed to be the same for the
two atoms. The electromagnetic field creation~(annihilation)
operator is $b_a^+$~($b_a$), while $S_+^{(n)}$~($S_-^{(n)}$) are the
spin raising~(lowering) operators for the $n^{\mbox{th}}$ atom. We
will define the pointing vector from $1$ to $2$ as $\br =\br_2
-\br_1$ and we will assume without loss of generality that $\br_1 +
\br_2 = 0$.

 The two 2LAs do not interact with each other directly
but only through the common electromagnetic field via the
interaction Hamiltonian
\begin{eqnarray}
\hat{H}_I = \hbar\sum_{a} g_a\left[ \hat{b}^{\dagger}_a (
e^{-i\bk\cdot\br/2} \hat{S}_{-}^{(1)} + e^{i\bk\cdot\br/2}
\hat{S}_{-}^{(2)} ) + b_a ( e^{i\bk\cdot\br/2} \hat{S}_{+}^{(1)} +
e^{-i\bk\cdot\br/2} \hat{S}_{+}^{(2)} ) \right]. \label{Hint}
\end{eqnarray}

While the coupling functions $g_a = g_{\bk \sigma}$ can be
completely general, its relevant form for the dipole coupling of the
electromagnetic field to the dipole of the atom ${\bf d}$ is
\begin{eqnarray}
g_a := g_{\bk \sigma} = \frac{\lambda}{\sqrt{V\omega_\bk}} {\bf
\hat{d}}\cdot {\bf e}_{\bk \sigma}, \label{ga}
\end{eqnarray}
where ${\bf e}_{\bk \sigma}$ are the polarization vectors, $ \lambda
{\bf \hat{d}}$ is the transition matrix element of the momentum
${\bf p}$ of the atomic electron (assumed reals) , ${\bf \hat{d}}$
being the corresponding unit vector; $V$ is the volume of space. We
have assumed that $ \lambda {\bf \hat{d}}$ is the same in both
atoms. The total Hamiltonian of the atom-field system is
\begin{equation}
\hat{H} = \hat{H}_0 + \hat{H}_I.
\end{equation}

The atom-field interaction Hamiltonian (\ref{Hint}) is
derived under the dipole and the rotating wave approximations (RWA)
(see Appendix A in Ref. \cite{AH}). The RWA keeps only the terms in
the interaction-picture Hamiltonian that correspond to resonant
coupling but ignores all the rapidly oscillating terms. For a single
qubit system the RWA is self-consistent. However, in the two-qubit
system since we keep terms in the Hamiltonian that vary in space as
$e^{i {\bf k} \cdot {\bf r}}$ the RWA is consistent only if
 $r << t$. This condition is satisfied in most realistic situations,
 but it fails in the formal limit $r \rightarrow \infty$.

 The two level approximation holds only  for photon frequencies that are
not much larger than $\w_o$. For this we introduce a high-frequency
cut-off $\epsilon^{-1}$ such that $\w_o \epsilon << 1$ \cite{AH}. In
what follows, the cut-off is implemented by inserting a factor $e^{-
\epsilon \omega_{\bf k}}$ in all integrations over momenta ${\bf k}$
that appear in the definition of the evolution functions. For $t, r
>> \epsilon^{-1}$, the cut-off dependence can be absorbed in a
renormalization of the basic parameters.

As the results in Section II B to II D do not depend on the explicit
form of the coupling constant $g_a$ (and they can also be applied to
non-optical systems with similar coupling) we leave the index $a$
completely general and only use the specific form (\ref{ga}) in Sec.
II E.

\subsection{Perturbative expansion and resummation}

We assume an initial factorized state at $t = 0$ for the combined
system of atoms+field of the form $|O \rangle \otimes | \psi
\rangle$, where $|O \rangle$ is the vacuum state of the EMF and
$|\psi \rangle$ is a vector on the Hilbert space of the two 2LA's.
To study the dynamics of this two 2LA-EMF system we seek the action
of the evolution operator $e^{-i\hat{H}t}$ on such vectors. The
calculations are simplified by using the resolvent expansion of the
Hamiltonian
\begin{eqnarray}
e^{-i\hat{H}t} = \frac{1}{2\pi} \int \frac{dE e^{-iE t}}{E - \hat{H}
+ i \eta}
\end{eqnarray}
which allows for a perturbation expansion
 \begin{eqnarray}
 (E - \hat{H})^{-1} = (E - \hat{H}_0)^{-1} + (E - \hat{H}_0)^{-1} \hat{H}_I (E - \hat{H}_0)^{-1}
\nonumber \\ + (E - \hat{H}_0)^{-1} \hat{H}_I (E - \hat{H}_0)^{-1}
\hat{H}_I (E - \hat{H}_0)^{-1} + \ldots \label{expansion}
\end{eqnarray}

Of relevance for the computation of the reduced density matrix of
the two qubits are matrix elements of the form $\langle z; i',j'|
 (E - \hat{H})^{-1}| O; i,j \rangle$, where $z$ refers to a coherent state of the EM
 field and $i, j$ take on the values of $0, 1$ corresponding to the
 ground state of a single qubit and to the excited state respectively.
We compute the matrix elements above through the perturbation
expansion (\ref{expansion}).
 A resummation of the perturbative series is possible, and it leads to an
exact expression for the matrix elements of the resolvent--see
Appendix A of Ref. \cite{ASH} for details. An inverse Fourier
transform yields the following values for the non-vanishing matrix
elements of the propagator.

\begin{eqnarray}
\langle z; 0, 0| e^{-i\hat{H}t}| O; 0, 1 \rangle &=& \sum_{a} e^{i
{\bf k}\cdot {\bf r}/2} z^*_{a} s_{a}(t) \label{00}\\
\langle z; 0, 1| e^{-i\hat{H}t}| O; 0, 1 \rangle &=& \int \frac{dE
e^{-iEt}}{2} \left[ \frac{1}{E - \w_o - \alpha(E) - \beta(E,r)}
\right. \nonumber \\ &+& \left.\frac{1}{E - \w_o - \alpha(E) +
\beta(E,r)} \right]
= : v_+(t) \label{iv+}\\
\langle z; 1, 1| e^{-i\hat{H}t}| O; 0, 1 \rangle &=& \int \frac{dE
e^{-iEt}}{2} \left[ \frac{1}{E - \w_o - \alpha(E) - \beta(E,r)}
\right. \nonumber \\ &-&
\left. \frac{1}{E - \w_o - \alpha(E) + \beta(E,r)} \right] =: v_-(t) \label{iv-}\\
 \langle z; 0, 0| e^{-i\hat{H}t}| O; 0, 0 \rangle &=& 1 \\
 \langle z; 1, 1| e^{-i\hat{H}t}| O; 1, 1 \rangle &=& \int \frac{dE
 e^{-iEt}}{E - 2 \w_o - 2 \alpha(E - \w_o) -f(E, r)} =: u(t) \label{iu}\\
 \langle z; 0, 0| e^{-i\hat{H}t}| O; 1, 1 \rangle &=& \int dE e^{-iEt}  \sum_{ab}
 \frac{ \hat{H}_{ab} z^*_a z^*_b }{ E - 2 \w_o - 2 \alpha(E - \w_o)
 -f(E, r)} \\
\left( \begin{array}{cc}  \langle z; 0, 1| e^{-i\hat{H}t}| O; 1, 1
\rangle
\\ \langle z; 1, 0| e^{-i\hat{H}t}| O; 1, 1 \rangle \end{array} \right)
&=& \sum_{a} z^*_a \left( \begin{array}{cc}
e^{-i\frac{{\bf k} \cdot {\bf r}}{2}} \nu_a(t) \\
e^{i\frac{{\bf k} \cdot {\bf r}}{2}} \nu'_a(t) \label{11}
\end{array} \right),
\end{eqnarray}
where,
\begin{eqnarray}
s_a(t)  &=&  \int \frac{dE e^{-iEt}}{(E-\w_o - \alpha(E) -
\beta(E,r) e^{i {\bf k} \cdot {\bf r}})(E -
\omega_a)} \label{sk}\\
\left( \begin{array}{cc} \nu_a(t) \\ \nu'_a(t)
\end{array} \right) &=& \int \frac{dE e^{-iEt}}{E - 2 \w_o} \sum_{b} (1 - L)_{ab} \left(
\begin{array}{cc} \frac{g_b}{E - \w_o - \omega_b} \\ \frac{g_{b}}{E - \w_o - \omega_b} \end{array} \right)
\label{nu} \\
 \alpha(E) :&=& \sum_a \frac{g_a^2}{E - \omega_a}
\label{al}
\\
\beta(E,r) :&=& \sum_a \frac{g_a^2}{E - \omega_a} e^{i{\bf k} \cdot
{\bf r} }. \label{be}
\end{eqnarray}

The matrix $L$ is
\begin{eqnarray}
L := \left( \begin{array}{cc} \Xi & \Theta \\ \bar{\Theta} &
\bar{\Xi}
\end{array} \right),
\end{eqnarray}
where
\begin{eqnarray}
\Xi_{ ab} = \frac{1}{E - \omega_o - \omega_a} \left( \alpha(E -
\omega_a) \delta_{ab} + g_{a} g_{b} (\frac{1}{E-2 \w_o} + \frac{e^{i
({\bf k} - {\bf k'}) \cdot {\bf r}}}{E - \omega_a - \omega_b}
)\right) \label{Xi}
\\
\Theta_{ab}  = \frac{1}{E - \omega_o - \omega_a} \left(
\beta(E-\omega_a, r) \delta_{ab} + g_{a} g_{b} (\frac{1}{E-2 \w_o} +
\frac{ 1}{E - \omega_a - \omega_b} )\right), \label{Theta}
\end{eqnarray}
and the overbar denotes complex conjugation.

The explicit definitions of the kernel $H_{ab}$ and of the function
$f$ can be found in \cite{ASH} but will not be needed for the issues
we are interested in here.

\subsection{The reduced density matrix}

We next compute the elements of the reduced density matrix for the
qubit system by integrating out the EM field degrees of freedom
\begin{eqnarray}
\rho^{ij}_{i'j'}(t) = \sum_{i_0,j_0,i'_0,j'_0}
\rho^{i_0,j_0}_{i'_0,j'_0} (0) \int [dz] [dz^*] \langle O;
i'_0j'_0|e^{i\hat{H}t}| z; i', j' \rangle \langle
z,i,j|e^{-i\hat{H}t} |O, i_0,j_0 \rangle, \label{rdm}
\end{eqnarray}
where $[dz]$ is the standard Gaussian integration measure for the
coherent states of the EM field.

Substituting Eqs. (\ref{00}-\ref{11}) into (\ref{rdm}) we obtain
through a tedious but straightforward calculation the elements of
the reduced density matrix

\begin{eqnarray} \rho^{I}_{I}(t) &=& \rho^{I}_{I}(0) |u|^2(t)
\label{1111}
\\
\rho^{11}_{01}(t) &=& \rho^{11}_{01}(0) u(t) v^*_+(t) +
\rho^{11}_{10}(0) u(t) v^*_-(t)
\\
\rho^{11}_{10}(t) &=& \rho^{11}_{10}(0) u(t) v^*_+(t) +
\rho^{11}_{01}(0) u(t) v^*_-(t)
\\
\rho^{I}_{00}(t) &=& \rho^{11}_{00} u(t)
\\
 \rho^{01}_{00}(t) &=& \rho^{01}_{00}(0) v_+(t) + \rho^{10}_{00}(0)
v_-(t) + \rho^{11}_{01}(0) \mu_1(t) + \rho^{11}_{10}(0) \mu_2(t)
\\
\rho^{10}_{00}(t) &=& \rho^{10}_{00}(0) v_+(t) + \rho^{10}_{00}(0)
v_-(t) + \rho^{11}_{01}(0) \mu^*_2(t) + \rho^{11}_{10}(0) \mu^*_1(t)
\\
\rho^{01}_{01}(t) &=& \rho^{01}_{01}(0) |v_+|^2(t) +
\rho^{01}_{10}(0) v_+(t) v^*_-(t) +  \rho^{10}_{10}(0) |v_-|^2(t)
\nonumber \\ &+& \rho^{10}_{01}(0)  v_-(t) v^*_+(t) +
\rho^{11}_{11}(0) \kappa_1(t)
\\
\rho^{01}_{10}(t) &=& \rho^{01}_{10}(0) |v_+|^2(t) +
\rho^{10}_{01}(0) |v_-|^2(t) +  \rho^{01}_{01}(0)  v_+(t)
v_-^*(t)\nonumber \\ & +& \rho^{10}_{10}(0) v_-(t) v^*_+(t) +
\rho^{11}_{11}(0) \kappa_2(t)
\\
\rho^{00}_{00}(t) &=& 1 - \rho^{11}_{11}(t) - \rho^{01}_{ 01}(t) -
\rho^{10}_{10}(t) \label{0000}
\end{eqnarray}
where
\begin{eqnarray}
\mu_1(t) &=& \sum_{a} g_a \nu_a(t) s^*_a(t) \label{defm1}
\\
\mu_2(t) &=& \sum_a g_a \nu_{a}(t) s^*_{a}(t)
e^{-i {\bf k} \cdot {\bf r}} \label{defm2}\\
\kappa_1(t) &=& \sum_{a}  |\nu_{a}|^2(t)  \label{defk1}\\
\kappa_2(t) &=& \sum_{a} \nu_{a}(t) \nu'^*_{a}(t) e^{- i {\bf k}
\cdot {\bf r}}, \label{defk2}
\end{eqnarray}
and the functions $u(t), v_{\pm}(t)$ were defined in Eqs.
(\ref{iu}), (\ref{iv+}) and (\ref{iv-}).

\subsection{Explicit forms for the evolution functions}

Eqs. (\ref{1111}-\ref{0000}) provide an {\em exact} expression for
the evolution of the reduced density matrix for the system of two
qubits interacting with an EM field in the vacuum state. The
evolution is determined by seven functions of time $u, v_{\pm},
\kappa_{1,2}, \mu_{1,2}$ defined above. To obtain the explicit forms
for these functions involves summing over all modes labeled by $a$,
namely, performing an integration over all momenta ${\bf k}$ and
summing over all polarizations.

We employ the following approximations:
\begin{enumerate}

\item We assume weak coupling ($\lambda^2 << 1$) and  ignore the
contribution of all processes that involve the exchange of two or
more photons between the two qubits.

\item We ignore branch cut terms that appear from the inverse
Fourier transform of the coefficients and keep only the contribution
of the dominant poles. One may prove that  the branch-cut term only
becomes significant at the very-long-time limit, well beyond the
time scales associated to decay.
\end{enumerate}

\subsubsection{The functions $u, v_{\pm}$}

With the approximation above, the contribution of the function $f$
drops out from the definition of $u$. Thus we obtain
\begin{eqnarray}
 u(t) &=& \int \frac{dE
 e^{-iEt}}{E - 2 \w_o - 2 \alpha(E - \w_o) } \label{uu}\\
 v_{\pm} &=& \int dE e^{-iEt}
\frac{1}{2}\left[ \frac{1}{E - \w_o - \alpha(E) - \beta(E,r)} \pm
\frac{1}{E - \w_o - \alpha(E) + \beta(E,r)} \right]. \label{vv}
\end{eqnarray}
For small $\lambda$ there is only one single pole and thus they
become
\begin{eqnarray}
u(t) &=& e^{ - 2 i \omega_o t - 2 \Gamma_0 t} \label{u}
\\
v_{\pm}(t) &=& \frac{e^{- i \omega_o t - \Gamma_0 t}}{2} \left( e^{
- i \sigma t - \Gamma_r t} \pm e^{i \sigma t + \Gamma_r t} \right).
\label{v+-}
\end{eqnarray}

 In the equations above, we renormalized the
frequency $\w_o$ by a constant divergent term  that arises from
$\alpha (E)$ (see \cite{AH}).  The parameters $\g_0, \Gamma_r$ and
$\sigma(r) $ are defined as
\begin{eqnarray}
\g_0 := - Im \, \alpha(\w_o) \\
-\sigma(r) + i \Gamma_r := \beta(\w_o, r).
\end{eqnarray}
For the coupling (\ref{ga}), and Eqs. (\ref{al}--\ref{be}) we find
\begin{eqnarray}
\g_0 &=& \frac{\lambda^2 \w_o}{3 \pi} \\
\Gamma_r &=& \frac{\lambda^2 }{2 \pi r}\left[( \sin \w_0 r +
\frac{\cos \w_o r}{\w_o r} - \frac{\sin \w_o r}{\w_0^2 r^2} )
\right. \nonumber \\
&-& \left.({\bf \hat{d}}\cdot {\bf \hat{r}})^2 (\sin \w_0 r + 3
\frac{\cos \w_o r}{\w_o r} - 3 \frac{\sin \w_o r}{\w_0^2 r^2})
\right],
\end{eqnarray}
where ${\bf \hat{r}}$ is the unit vector in the direction of
$\bf{r}$.

The term $\s(r)$ is a frequency shift caused by the vacuum
fluctuations. It breaks the degeneracy of the two-qubit system and
generates an effective dipole coupling between the qubits. The
constant $\g_0$ corresponds to the rate of emission from individual
qubits; it is the same as that from a single qubit interacting with
the electromagnetic field. The function $\Gamma_r$ is specific to
the two-qubit system. It arises from  Feynman diagrams that involve
an exchange of photons between the qubits.  As $r \rightarrow 0$,
$\Gamma_r \rightarrow
 \g_0$ and as $r \rightarrow \infty$, $\Gamma_r \rightarrow 0$. Note
 that the ratio $\g_r/\g_0$, while smaller than unity, is of
 the order of unity as long as $r$ is not much larger than
 $\omega_o^{-1}$.

\subsubsection{The functions $\kappa_{1,2}(t)$}
We first compute the functions $\nu_a, \nu'_a$ of Eq. (\ref{nu})
keeping terms up to second loop order
\begin{eqnarray}
\left( \begin{array}{cc} \nu_a(t) \\ \nu'_a(t)
\end{array} \right) &=& \int \frac{dE e^{-iEt}}{E - 2 \omega_o} \sum_b  \frac{g_b}
{E - \omega_o - \omega_b}\left(
\begin{array}{cc} \delta_{ab} + \Xi_{ ab} + \Theta_{ ab} \\\delta_{ ab} +
 \bar{\Xi}_{ ab} + \bar{\Theta}_{ ab}\\  \end{array}
\right),
\end{eqnarray}
where $\Xi$ and $\Theta$ are given by Eqs. (\ref{Xi}) and
(\ref{Theta}). To leading order in $\lambda^2$ we obtain
\begin{eqnarray}
\nu_{a}(t) = \nu'_{a}(t) = g_{a} \int \frac{dE e^{-iEt}}{[E - 2
\omega_o - 2 \alpha (E - \omega_o)][E - \omega_o - \omega_a -
\alpha(E - \omega_a) - \beta(E - \omega_a, r))]}. \label{nnn}
\end{eqnarray}
Keeping only the pole terms from the integral in (\ref{nnn}) we
obtain
\begin{eqnarray}
\nu_a(t) = g_a \frac{e^{ - i \w_o t - \Gamma_0 t}}{\omega_o -
\omega_a - \sigma -i \Gamma_0 + i \Gamma_r} \left( e^{- i \omega_0 t
- \Gamma_0t} - e^{ -i \omega_a t - i \sigma t - \Gamma_r t} \right).
\end{eqnarray}
We then substitute the expression above for $\nu_a$ into Eqs.
(\ref{defk1}) and (\ref{defk2}). Using the values of Eq. (\ref{ga})
for $g_a$, and summing over the polarizations and angular variables,
we obtain
\begin{eqnarray}
\kappa_1(t) &=& \frac{\lambda^2}{3 \pi^2} e^{-2 \Gamma_0t}
\int_0^{\infty} k dk \, \frac{e^{-2 \Gamma_0 t} + e^{ - 2\Gamma_r t}
- 2 e^{ - (\Gamma_0 + \Gamma_r) t} \cos [ (\omega_o - k -
\sigma)t]}{(k - \omega_o +
\sigma)^2 + (\Gamma_0 - \Gamma_r)^2} \label{k1}\\
\kappa_2(t) &=& \frac{\lambda^2}{2 \pi^2 r} e^{-2 \Gamma_0t}
\int_0^{\infty} dk \,  \frac{e^{-2 \Gamma_0 t} + e^{ - 2\Gamma_r t}
- 2 e^{ - (\Gamma_0 + \Gamma_r) t} \cos [ (\omega_o - k -
\sigma)t]}{(k - \omega_o + \sigma)^2 + (\Gamma_0 - \Gamma_r)^2}
\nonumber \\
&\times& \left[( \sin k r + \frac{\cos k r}{k r} - \frac{\sin k
r}{k^2 r^2} )-({\bf \hat{d}}\cdot {\bf \hat{r}})^2 (\sin k r + 3
\frac{\cos k r}{k r} - 3 \frac{\sin k r}{k^2 r^2}) \right]
\label{k2}
\end{eqnarray}

To compute these functions we use an approximation scheme of
replacing the Lorentzian in (\ref{k1}-\ref{k2}) by a delta function,
which is valid for time-scales $ t >> \w_o^{-1}$, i.e.,
\begin{eqnarray}
\frac{1}{(k - \w_o + \sigma)^2 + (\Gamma_0 - \Gamma_r)^2} \simeq
\frac{\pi}{\Gamma_0 - \Gamma_r} \delta (k - \w_o + \sigma) + O\left[
\left(\frac{\Gamma_0 - \Gamma_r}{\w_o}\right)^2\right].
\label{approximation}
\end{eqnarray}
To leading order in $\lambda^2$ we obtain the simple expressions
\begin{eqnarray}
\kappa_1(t) = \Gamma_0 \kappa(t) \label{kappa1} \\
\kappa_2(t) = \Gamma_r \kappa(t) \label{kappa2}
\end{eqnarray}
where
\begin{eqnarray}
\kappa(t) \simeq \frac{1}{\Gamma_0 - \Gamma_r} e^{-2 \Gamma_0t}
(e^{-\Gamma_0 t} - e^{- \Gamma_rt})^2  \label{kappa}.
\end{eqnarray}



\section{Disentanglement of two qubits}

In this section, we employ the results obtained above to study the
evolution of the two qubits initially in an entangled state. We
shall focus on the process of disentanglement induced by their
interaction with the field.

\subsection{Class A states: Initial superposition of $|00 \rangle$ and $|11\rangle$}

We first examine the class of initial states we call Class A of the
following type
\begin{eqnarray}
| \psi_o \rangle = \sqrt{1-p} |00\rangle + \sqrt{p} |11 \rangle,
\label{gggg}
\end{eqnarray}
where $0 \leq p \leq 1$. Recall our definition $|I \rangle = |11
\rangle$ and  $|O \rangle = |00 \rangle$. From Eqs. (\ref{1111} -
\ref{0000}) we obtain
\begin{eqnarray}
\hat{\rho}(t) = p^2 e^{-4 \g_0 t} |I \rangle \langle I| + e^{-2 \g_0
t} \sqrt{p(1-p)} (  e^{2 i \omega_o t} |I\rangle \langle O| +
 e^{-2 i \omega_0 t}  \, |O \rangle \langle I|) \nonumber
\\ + p [ \kappa_1(t)
- \kappa_2(t)] |- \rangle \langle -| + p [\kappa_1 (t) +
\kappa_2(t)] |+ \rangle \langle +| + [1 - p^2 e^{-4 \g_0 t} - 2p
\kappa_1(t)] |O \rangle \langle O|, \label{roa}
\end{eqnarray}
where the functions $\kappa_1(t)$ and $\kappa_2(t)$ are given by
Eqs. (\ref{kappa1}--\ref{kappa2}).

These results are quite different from those reported in Ref.
\cite{FicTan06}, which were obtained under the Born-Markov
approximation. While the $|I \rangle \langle I|$ and $|I\rangle
\langle O|$ terms are essentially the same, the $|- \rangle \langle
-|$ and $|+ \rangle \langle +|$ ones are not, as they involve
non-diagonal elements of the density matrix propagator. For
comparison, we reproduce here the explicit form of these matrix
elements in our calculation
\begin{eqnarray}
\rho^+_+ &=& p \frac{\Gamma_0 + \Gamma_r}{\Gamma_0 - \Gamma_r} e^{-2
\Gamma_0t} (e^{-\Gamma_0 t} - e^{- \Gamma_rt})^2 \label{rho++} \\
\rho^-_- &=& p e^{-2 \Gamma_0t} (e^{-\Gamma_0 t} - e^{-
\Gamma_rt})^2, \label{rho+-}
\end{eqnarray}
and in that of Ref. \cite{FicTan06}
 (translated into our notation):
\begin{eqnarray}
\rho^+_+ (t) &=& p \frac{\Gamma_0 + \Gamma_r}{\Gamma_0 - \Gamma_r}
e^{-2 \Gamma_0 t} ( e^{-2 \Gamma_rt} - e^{- 2 \Gamma_0 t}) \label{FTt} \\
\rho^-_-(t) &=&  p \frac{\Gamma_0 - \Gamma_r}{\Gamma_0 + \Gamma_r}
e^{-2 \Gamma_0 t} ( e^{2 \Gamma_rt} - e^{- 2 \Gamma_0 t}).
\label{FTs}
\end{eqnarray}
For large values of $r$, $\Gamma_r << \g_0$ the expressions
above coincide. However, for smaller values of $r$ their difference
is significant. We note that even though $r > a_B$ the regime $\w_o
r << 1$ is physically meaningful, as long as $\w_o a_B <<1$. In this
regime the difference between (\ref{rho+-}) and (\ref{FTs}) is substantial:
$\g_0 \simeq \Gamma_r$ and at times $\g_0 t \sim 1$ we
obtain $(\g_0 - \Gamma_r)t <<1$. According to the Markovian results
of \cite{FicTan06}, in this regime the $|+ \rangle \langle +|$ term
is of order $O(\lambda^0)$ and hence comparable in size to the other
terms appearing in the evolution of the density matrix. However,
according to our results, which are based on the full non-Markovian
dynamics, the $|+ \rangle \langle +|$ term is of order $\frac{\g_0 -
\Gamma_r}{\g_0}$ and hence much smaller.

In general, for $\w_0 r << 1$, we find that  the $|- \rangle \langle
-|$ and $|+ \rangle \langle +|$ terms contribute little to the
evolution of the reduced density matrix and they can be ignored.
Since these terms are responsible for the sudden death and
subsequent revival of entanglement studied in \cite{FicTan06}, we
conclude that these effects are absent in this regime.

\subsection{Bell states}

For the case of an initial Bell state \cite{Bell} $|\pm \rangle$ for
the two qubits we find
\begin{eqnarray}
\hat{\rho}(t) = e^{ -2 [\g_0 \pm \Gamma_r]t} |\pm \rangle \langle
\pm| + \left(1 -e^{ -2 [\g_0 \pm \Gamma_r]t}\right)|O \rangle
\langle O |
\end{eqnarray}
We see that the symmetric  $|+ \rangle$ state decays super-radiantly
with decay rate $\g_0 + \Gamma_r$ and the anti-symmetric state $| -
\rangle$ decays sub-radiantly with the rate $\g_0 - \Gamma_r$. The
results agree qualitatively with those obtained in Ref.
\cite{FicTan06} under the Born-Markov approximation.

\subsection{Non-Markovian Features}

Let us try to understand these results with some discussions.

\subsubsection{Differences from Born-Markov}

For an initial Class A state  (\ref{gggg}),  the $|I \rangle$
component decays to the vacuum, but it also evolves into a linear
combination of antisymmetric $|- \rangle $  and symmetric $|+
\rangle $  Bell states. However, if the qubits are close together
the evolution to Bell states is suppressed. This behavior is
qualitatively different from that of Ref. \cite{FicTan06}, which was
obtained through the Born-Markov approximation. The corresponding
terms differ substantially at the physically relevant time-scales.
As a consequence, we find that there is neither sudden death nor
revival of entanglement in this regime.

In order to explain this difference we note that the Born-Markov
method involves two approximations: i) that the back-action of the
field on the atoms is negligible and ii) that all memory effects in
the system are insignificant. When the qubits are found within a
distance much smaller than their characteristic wavelengths, it is not possible to ignore back-action.
The virtual photons  exchanged by the qubits (at a rate given by
$\Gamma_r$) substantially alter the state of the electromagnetic
field.

On the other hand, the effect of virtual photons exchange between
qubits drops off quickly at large separations $r$  -- the two qubits
decay almost independently one of the other. Hence, the Born-Markov
approximation -- reliable for the case of two separate qubits each
interacting with an individual field -- also gives reasonable
results for the two qubits interacting with a common field. In this
regime sudden death is possible, but not revival of entanglement. In
this sense, our results effectively reduces to those of Ref.
\cite{YuEbePRL}: when the distance between the qubits is much larger
than any characteristic correlation length scale of the system it
behaves as if the two qubits are found in different reservoirs.


\subsubsection{The origin of the non-Markovian behavior}


From Eqs. (\ref{1111}--\ref{0000}) and (\ref{u}, \ref{v+-}) we note
that the diagonal terms of the reduced density matrix propagator all
decay exponentially, which is a characteristic  sign of Markovian
behavior.  Hence, as far as this part of the evolution is concerned,
our results are fully compatible with the Markovian predictions.

However, the behavior of the non-diagonal terms in the reduced
density matrix propagator is different. Eqs. (\ref{1111}-\ref{0000})
show that the only non-zero such terms are ones that describe the
effect of successive decays, for example the $|11 \rangle $ state
first decaying into $|- \rangle$ and then $|- \rangle $ decaying
into the ground state $|00 \rangle$. Hence, the $\rho^-_-(t)$ term
consists of one component that contains the remaining of the $|-
\rangle \langle -|$ part of the initial state and another component
that incorporates the decay of the $|11 \rangle \langle 11|$ part of
the initial state towards the state $|- \rangle$. In our
calculation, the latter term is encoded into the functions
$\kappa_{1,2}(t)$, which are obtained by squaring the amplitudes
$\nu_{\bf k}(t)$ as in Eqs. (\ref{defk1}--\ref{defk2}). The
amplitudes $\nu_{\bf k}(t)$ are obtained from the summation of two
Feynman diagrams --see Eq. (\ref{nnn}). The structure of the poles
in Eq. (\ref{nnn}) reveals that the first Feynman diagram describes
the decays of the $|11 \rangle$ state, while the second one
corresponds to processes involving the $|01 \rangle$ and $|10
\rangle$ states.

The key point is that the evolution functions   $\kappa_{1,2}(t)$
are squares of the corresponding amplitudes, and for this reason
they contain {\em interference terms} between the two types of
processes that contribute to $\nu_{\bf k}(t)$. Major difference of
our non-Markovian results stems from the constituents of the
off-diagonal terms. In the Markov approximation these term involve
summation (subtraction) of probabilities rather than of amplitudes;
hence, it ignores the interference between these two processes.

\section{Decoherence of Two qubits}

We now turn our attention to the decoherence issue. We want to
compare the rate of decoherence of the two qubit system  by the EMF
with that of one qubit alone in order to understand how the
decoherence of one qubit is affected by the presence of another. We
assume that the initial state of the two 2LA-EMF system is of the
form
\begin{eqnarray}
\left(\sqrt{p} |1 \rangle + \sqrt{1-p} |0 \rangle \right) \otimes |0
\rangle =  \sqrt{p} |10 \rangle + \sqrt{1-p} |00 \rangle.
\end{eqnarray}
where the first qubit is prepared initially in a superposition of
the $|0 \rangle$ and $|1 \rangle$ states, and the second qubit lies
on the ground state. From Eqs. ( \ref{1111}--\ref{0000}), we obtain
the density matrix of the combined qubit system
\begin{eqnarray}
\hat{\rho}(t) = p \left( |v_+|^2 |10\rangle \langle 10| + |v_-|^2
|01 \rangle \langle 01 | + v_-^* v_+ |01 \rangle \langle 10 | +
v_- v_+^* |10 \rangle \langle 01| \right) \nonumber \\
+ \sqrt{p(1-p)} \left( v_+^* |10 \rangle \langle 00| + v_-^* |01
\rangle \langle 00| + v_+ |00 \rangle \langle 10| + v_- |00 \rangle
\langle 01| \right) \nonumber \\
+ \left(1 - p (|v_+|^2 + |v_-|^2) \right) |00 \rangle \langle 00|,
\end{eqnarray}
where the functions $v_{\pm}(t)$ are given by Eq. (\ref{v+-}).

The two qubits become entangled through their interaction via the EM
field. To study the  decoherence in the first qubit, we trace out
the degrees of freedom of the second one, thus constructing the
reduced density matrix $\hat{\tilde{\rho}}_1$
\begin{eqnarray}
\hat{\tilde{\rho}}_1(t) = p |v_+|^2 |1 \rangle \langle 1| +
\sqrt{p(1-p} \left( v_+ |0\rangle \langle 1| + v_+^* |1 \rangle 0|
\right) + \left( 1 - p|v_+|^2 \right) |0 \rangle \langle 0|.
\end{eqnarray}

At large interqubit separations $\Gamma_r = 0 = \sigma(r)$, whence
$v_+ \simeq e^{- i \w_o t - \g_0 t}$, the off-diagonal elements
decay within a characteristic time-scale of order $\g_0^{-1}$. These
results coincide with those for the single qubit case--see Refs.
\cite{AH, SADH}. However, for smaller values of $r$, the results are
substantially different. The entanglement with the second qubit
leads to a departure from pure exponential decay. In particular, for
$\w_or << 1$, $\Gamma_r \simeq \g_0$. This implies for times longer
than $\g_0^{-1}$ a substantial fraction of the off-diagonal elements
persists. This decays eventually to zero within a time-scale of
order $[\g_0 - \Gamma_r]^{-1} >> \g_0^{-1}$. Hence, the qubit
preserves its coherence longer when there is another quibit close
by.

The reduced density matrix of the second qubit is
\begin{eqnarray}
\hat{\tilde{\rho}}_2(t) = p |v_-|^2 |1 \rangle \langle 1| +
\sqrt{p(1-p)} \left( v_- |0\rangle \langle 1| + v_-^* |1 \rangle 0|
\right) + \left( 1 - p |v_-|^2 \right) |0 \rangle \langle 0|.
\end{eqnarray}
Note that at small inter-qubit separations the asymptotic behavior (
for $\g_0 t >> 1$) of $\hat{\tilde{\rho}}_1(t)$ is identical to that
of $\hat{\tilde{\rho}}_2(t)$. The second qubit (even though
initially in its ground state) develops a persistent
quantum coherence as a result of the interaction with the first one.

It is also of interest to consider thermal field environment. Shresta et al \cite{SADH} have studied the decoherence of one qubit in a finite temperature EMF.  Yu and Eberly \cite{nqbthf} have considered the separability of the joint state of a collection of two-level systems at finite temperature. They conclude that since only separable states are found in the neighborhood of their thermal equilibrium state unimpeded thermal decoherence will destroy any initially arranged entanglement in a finite time.\\

\noindent{\bf Acknowledgement} This is a synopsis of the unpublished
paper \cite{ASH} supported by grants from the NSF (PHY-0426696), NIST,  NSA-LPS and by a Pythagoras II grant (EPEAEK).

\end{document}